\documentclass[journal,10pt,twocolumn]{IEEEtran}
\usepackage[a4paper,margin={1.61 cm,1.61 cm,1.61 cm,1.61 cm}]{geometry}
\usepackage{graphics}
\usepackage{epsfig}
\usepackage{subfigure}
\usepackage{amsmath}
\usepackage{cite}
\usepackage{color}
\usepackage{tabularx}

\usepackage{amssymb}
\usepackage{amsfonts} 
\usepackage{booktabs}
\usepackage{tabularx}
\usepackage{makecell}
\usepackage[
            CJKbookmarks=true,
            bookmarksnumbered=true,
            bookmarksopen=true,
            colorlinks=true,
            citecolor=blue,
            linkcolor=red,
            anchorcolor=green,
            urlcolor=blue
            ]{hyperref}

\IEEEoverridecommandlockouts                              

\title{Privacy-Preserving State Estimation in the Presence of Eavesdroppers: A Survey}

\author{
Xinhao~Yan,
Guanzhong Zhou,
Daniel~E.~Quevedo,
Carlos~Murguia,
Bo~Chen,
Hailong~Huang
\thanks{Xinhao Yan, Guanzhong Zhou, and Hailong Huang are with the Department of Aeronautical and Aviation Engineering, The Hong Kong Polytechnic University, Hong Kong (email: xin-hao-shawn.yan@connect.polyu.hk; guanzhong.zhou@connect.polyu.hk; hailong.huang@polyu.edu.hk).
}
\thanks{Daniel Quevedo and Carlos Murguia are with the School of Electrical Engineering and Robotics, Queensland University of Technology, Brisbane, QLD 4000, Australia (emails: daniel.quevedo@qut.edu.au, carlos.murguia@qut.edu.au).
}
\thanks{Bo Chen is with the Department of Automation, Zhejiang University of Technology, Hangzhou 310023, China (email: bchen@aliyun.com).
}
\thanks{\emph{Corresponding author: Hailong Huang.}}
}

\begin{document}

\markboth{}
{Yan \MakeLowercase{\textit{et al.}}: 
Privacy-Preserving State Estimation in the Presence of Eavesdroppers: A Survey 
}

\maketitle

\begin{abstract}
Networked systems are increasingly the target of cyberattacks that exploit vulnerabilities within digital communications, embedded hardware, and software. Arguably, the simplest class of attacks -- and often the first type before launching destructive integrity attacks -- are eavesdropping attacks, which aim to infer information by collecting system data and exploiting it for malicious purposes. A key technology of networked systems is state estimation, which leverages sensing and actuation data and first-principles models to enable trajectory planning, real-time monitoring, and control. However, state estimation can also be exploited by eavesdroppers to identify models and reconstruct states with the aim of, e.g., launching integrity (stealthy) attacks and inferring sensitive information. It is therefore crucial to protect disclosed system data to avoid an accurate state estimation by eavesdroppers. This survey presents a comprehensive review of existing literature on privacy-preserving state estimation methods, while also identifying potential limitations and research gaps. Our primary focus revolves around three types of methods: cryptography, data perturbation, and transmission scheduling, with particular emphasis on Kalman-like filters. Within these categories, we delve into the concepts of homomorphic encryption and differential privacy, which have been extensively investigated in recent years in the context of privacy-preserving state estimation. Finally, we shed light on several technical and fundamental challenges surrounding current methods and propose potential directions for future research.
\end{abstract}

\def\abstractname{Note to Practitioners}
\begin{abstract}
With the increasing openness and anonymization of the networked estimation systems, privacy concerns require to be paid more attention.
The essence of the encryption approaches is to seek certain tradeoffs among various performance indexes.
For instance, cryptography is suitable for high-performance processors, because it needs sufficient computation resources to generate and operate complicated secret keys.
On the other hand, perturbation-based encryption can be realized fast, but its adverse impact on the legitimate systems should be limited not to disrupt the desired operations.
In conclusion, the selection of these methods is based on the practical demands.
Besides, the discussed methods are all based on the general state-space model, thus they can represent most real-world dynamics.
It means that these approaches can be easily deployed to the practical engineering systems to effectively guarantee their privacy.
\end{abstract}


\begin{table}
  \caption{Notations \& Acronyms}
  \label{Table_Notations}
  \centering
  \begin{tabular*}{8cm}{ll}
  \hline
  $\triangleq$              &  define  \\
  $\mathbb{Z}$              &  set of integers  \\
  $\mathbb{Z}_{+}$           &  set of positive integers  \\
  $\mathbb{R}$              &  set of real numbers  \\
  $\mathbb{R}^{n}$          &  set of $n$-dimensional real vectors  \\
  $\mathbb{R}^{n\times m}$  &  set of $n\times m$ real matrices  \\
  $\mathbb{E}\{\cdot\}$     &  mathematical expectation   \\
  $\mathbb{P}\{\cdot\}$     &  probability function   \\
  $\mathrm{Tr}\{\cdot\}$    &  trace of a matrix  \\
  $X>(<)0$                  &  positive-definite (negative-definite)  \\
  $X\geq(\leq)0$            &  non-negative definite (non-positive definite)  \\
  $d'$                      &  adjacent variable of $d$  \\
  $A^{\mathrm{T}}$          &  transpose of real matrix $A$  \\
  $\mathrm{A}^{\dag}$       &  Hermitian transpose of complex matrix $\mathrm{A}$  \\
  $I$                       &  identity matrix with certain dimension  \\
  AVC                       &  aerial vehicle carrier \\
  CI                        &  covariance intersection \\
  CSE                       &  centralized state estimation \\
  DFE                       &  distributed fusion estimation \\
  DP                        &  differential privacy \\
  DSE                       &  distributed state estimation \\
  EKF                       &  extended Kalman filter  \\
  FC                        &  fusion center \\
  FHE                       &  fully homomorphic encryption \\
  HE                        &  homomorphic encryption \\
  IF                        &  information filter \\
  iff                       &  if and only if \\
  IO                        &  interval observer \\
  ITS                       &  intelligent transportation system \\
  i.i.d.                    &  independently and identically distributed \\
  KF                        &  Kalman filter \\
  LO                        &  Luenberger observer \\
  LQG                       &  linear quadratic Gaussian  \\
  LS                        &  least square \\
  LSE                       &  local state estimate \\
  MHE                       &  moving horizon estimator  \\
  MMSE                      &  minimum mean square error \\
  MSE                       &  mean square error \\
  PHE                       &  partially homomorphic encryption \\
  SE                        &  squared error \\
  SMPC                      &  secure multi-party computation \\
  SNR                       &  signal-to-noise ratio \\
  UKF                       &  unscented Kalman filter  \\
  WGN                       &  white Gaussian noise \\
  WLS                       &  weighted least square \\
  \hline
  \end{tabular*}
\end{table}

\begin{figure*}[ht]         
    \centering
    \includegraphics[width=15cm]{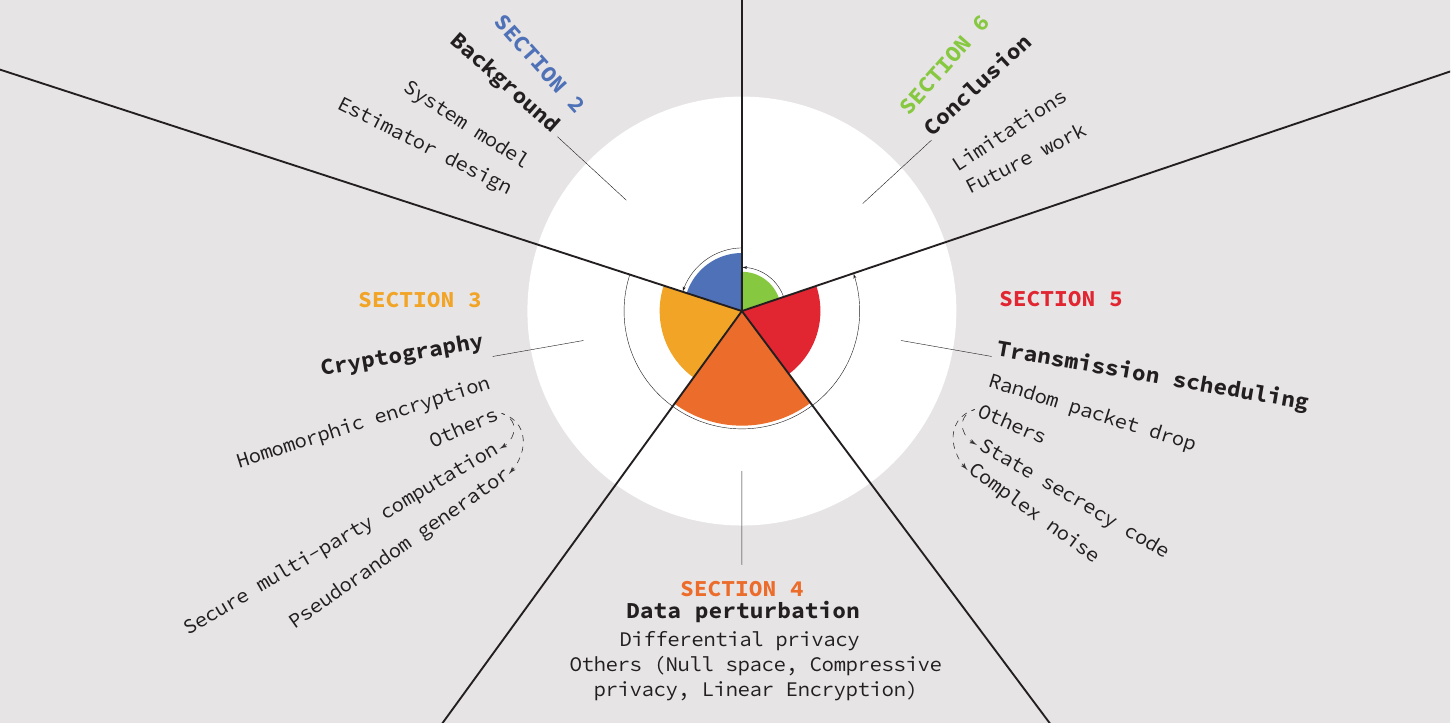}
    \caption{The organization of this paper. 
    Section 2 gives the general state-space models and corresponding estimator. 
    Sections 3, 4, and 5 respectively discuss cryptography, data perturbation, and transmission scheduling. 
    Finally, Section 6 summarizes the benefits and limitations, and provides some possible future work.}
\label{Fig_Organization}
\end{figure*}

\section{Introduction} 
State estimation refers to the process of inferring the value of a system state variable from a limited number of measurement data \cite{Shalom_Est}. 
In recent years, state estimation has gained considerable attention, because it can provide a valuable understanding of the dynamic behavior and condition. 
According to the existing literature, there are two basic state estimation structures, named centralized state estimation (CSE) \cite{Paul_Centr,Tsiamis_Centr} and distributed state estimation (DSE) \cite{Rego_Distr,Primadianto_Distr,Ge_Distr,An_Distr,Wu_Distr,Doostmohammadian_Distr,Liu_Distr}.
In the CSE structure, the information from all subsystems will be sent to an estimation center and the global state will be estimated together. 
On the other hand, in the DSE structure, the state of each subsystem will be estimated individually with its local measurements and the exchanged information from other subsystems. 
Moreover, in the multi-sensor field, there exist two fusion structures \cite{XiaorongLi_Fusion,Smith_Fusion,Li_Fusion}, namely centralized fusion estimation (CFE) \cite{Qian_Centr,Zhou_Centr,Pham_Centr} and distributed fusion estimation (DFE) \cite{Sun_Distr,Sun_Distr2,Sun_Distr3,Li_Distr,BoChen_Distr_TAC,Michieletto_Distr}. 
Within the CFE structure, all the raw measurements will be sent to a fusion center (FC) directly without any processing. 
By contrast, the measurements will be firstly processed at sensor sides in the DFE structure, generally into local state estimates (LSE), and then the processed data will be transmitted to the FC. 

Thanks to the low wiring complexity and cost, networks have been exploited to connect the spatially deployed parties in the estimation systems \cite{BoChen_Network_TAES}. 
However, with the openness of networks, anonymity of users, and remote connection among distributed parties, the transmissions via networks are vulnerable to a range of cyberattacks \cite{Ding_Attack,Ding_Attack2,Pajic_Attack,Sedjelmaci_Attack}. 
Denial-of-service (DoS) attacks aim to make real-time data unavailable to its intended users by temporarily or indefinitely disrupting the services of the sensor connected to the network \cite{Pirayesh_Jamming,Agrawal_DoS,Zhang_DoS,Wang_DoS,Qi_DoS,Wang2_DoS,Yang_DoS,Jiang_DoS,Shi_DoS}, and false data injection (FDI) attacks insert the wrong or useless information into the communication channels to mislead the system or disrupt its functioning \cite{Zhu_Deception,Li_Deception,Li_Replay,Liu_Replay,Deng_FDI,Miao_FDI,Gao_FDI,An_FDI,Zhu_FDI,Hu_FDI}. 
Note that most cyberattacks will cause direct damage to systems while eavesdropping attacks will not. 
Eavesdropping is a type of passive attack that only silently overhears the transmission of systems without any modification \cite{Kapetanovic_Eaves,Wang_Eaves,Conti_Eaves,Zheng_Eaves,Yuan_Eaves}. 
That is not to say that eavesdropping will not harm the system, because the wiretapped data can be further used to devise other more aggressive cyberattacks. 
Meanwhile, the data itself is very valuable, and the leakage of private information will cause unintended consequences. 

For countering malicious third-parties, many privacy-preserving methods have been discussed, for instance, physical-layer encryption approaches \cite{Aldaghri_Physical,Xu_Physical}. 
The legitimate systems can employ some protocols such as authentication \cite{Zhang_Authentication} to prevent eavesdroppers from accessing the communication channels, thereby ensuring complete privacy through the physical layer. 
Note that the universality of the physical-layer encryption method may be relatively poor, as it is designed to protect data at the physical layer, and may not be effective in securing data at rest or during processing. 
Furthermore, constant updates and maintenance are required to ensure they are equipped with the latest security schemes. 
Hence, this paper mainly concerns the data-aimed methods that are more general and can be applicable to most systems. 
In general, the main data-based methods consist of cryptography \cite{Goldreich_Crypt,Goldreich_Crypt2,Alexandru_Crypt,Sharma_Crypt,Shim_Crypt,Shim_Crypt2}, data perturbation \cite{Dwork_DP,Dwork_DP2,Dwork_DP2014,Kailkhura_Survey,Hassan_DP,Comas_DP,Ye_DP,Wu_DP,Ekenstedt_DP}, and transmission scheduling \cite{Quevedo_Drop,Leong_Drop_TAC,Abbas_Drop}. 
Each method has its own benefits and drawbacks, for instance, the computation amount of cryptography is very high, but the accuracy of data can be retained. 
On the contrary, the computation burden for data perturbation is low, while the perturbation will affect the originally good performance of legitimate users. There are recent perturbation methods that do not lead to performance degradation \cite{Carlos1}. They exploit ideas from system immersion theory and homomorphic encryption. The detailed description of these methods will be given in the corresponding sections.

\begin{figure*}[ht]         
    \centering
    \includegraphics[width=18cm]{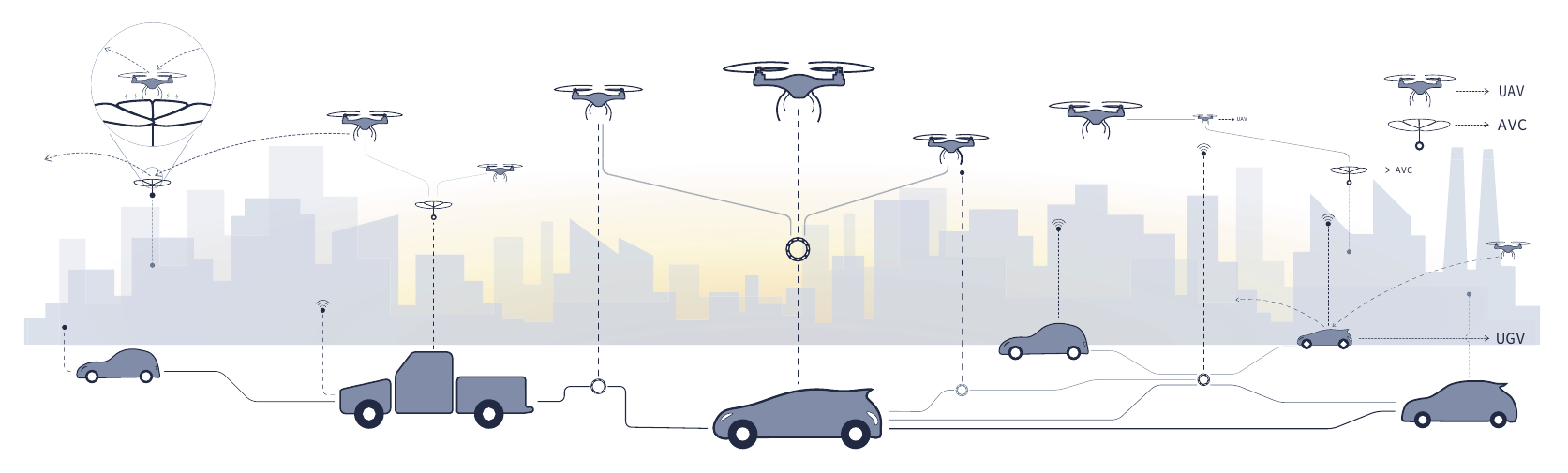}
    \caption{An example of state estimation. 
    The UGV cluster on the ground and the UAV cluster in the sky establish their own communication networks. 
    At the same time, the UAVs and UGVs also establish communication mechanisms among them for some collaborative missions. 
    Individuals in the communication network exchange information such as positions, speeds, accelerations, remaining powers, and task processes among each other, so as to estimate the state of themselves and the targets. 
    Meanwhile, the UAVs can also estimate the states of AVCs or UGVs for landing, and then they can complete tasks such as charging and parcel delivery. 
    }
\label{Fig_ITS}
\end{figure*}

To the best of the authors’ knowledge, there is currently a scarcity of surveys that thoroughly explore the latest advancements in protecting the privacy of estimators in the presence of eavesdroppers. 
Therefore, it is of high significance to conduct a comprehensive survey and offer detailed analyses of recent results in privacy-preserving state estimation methods. 
Additionally, we provide updated assessments regarding the limitations of existing results, followed by proposing potential research directions for future exploration. 

The remainder of this paper is organized as follows. 
First, the system model and estimator structure are revisited in Section II. 
Then, the cryptography, data perturbation, and transmission scheduling methods are respectively discussed in Sections III, IV, and V. 
In these sections, we first introduce the fundamentals of these methods and discuss relevant works. 
Subsequently, we focus on the discussions on estimator designs that leverage these methods, encompassing the aforementioned centralized, distributed, fusion estimation structures. 
Finally, Section VI summarizes the limitations of the methods, and gives some insights and possible research areas in the future. 
The complete organization of this paper is shown in Fig. \ref{Fig_Organization}, and the notations and acronyms frequently used throughout the paper are summarized in Table \ref{Table_Notations}.

\section{Privacy-Preserving State Estimation} 
A general description where the system states (physical variables of interest) evolve according to a set of nonlinear difference equations can be written as follows: 
\begin{eqnarray}                              
    \label{x_nonlinear}
    &&x(t+1)=f(x(t),t)+w(t), \\
    \label{y_nonlinear}
    &&y(t)=g(x(t),t)+v(t).
\end{eqnarray}
where $t \in \mathbb{Z}$ is a discrete-time variable, $x(t)\in\mathbb{R}^{n}$ represents the system state, $y(t)\in\mathbb{R}^{m}$ is the measurement or system output, and $f(x(t),t)\in\mathbb{R}^{n}$ and $g(x(t),t)\in\mathbb{R}^{m}$ are nonlinear time-varying vector functions that model the state and measurement evolution in time. Vector $w(t)\in\mathbb{R}^{n}$ denotes control actions, references, and process noise, and $v(t)\in\mathbb{R}^{m}$ models measurement noise caused by, for instance, inaccurate modeling and unknown physical properties.

Because, in general, models of the form \eqref{x_nonlinear}-\eqref{y_nonlinear} are highly complex and hard to work with, simplified models are often used in the literature. These models are obtained by linearizing nonlinear equations around equilibrium positions, operating points, or tracked trajectories. Linearized models can be written as follows: 
\begin{eqnarray}                              
    \label{x_linear}
    &&x(t+1)=Ax(t)+w(t), \\
    \label{y_linear}
    &&y(t)=Cx(t)+v(t),
\end{eqnarray}
where $A\in\mathbb{R}^{n\times n}$ denotes the state transition matrix determined by the system dynamics and $C\in\mathbb{R}^{m\times n}$ denotes the measurement matrix determined by the sensing characteristics. 
Most studies in the literature about privacy-preserving state estimation focus on time-invariant linear state-space models of the form \eqref{x_linear}-\eqref{y_linear}. 

For distributed systems, the state transition of each subsystem can be written as $x_{i}(t+1)=A_{i}x_{i}(t)+\sum_{j=1}^{M}A_{ij}x_{j}(t)+w_{i}(t)$. 
This means that each subsystem will be affected by the neighboring dynamics in addition to its own recursion. 
It is a common expression for platooning systems, for instance, the velocity and inter-vehicle distances of a vehicle in a platoon are regulated by the positions, velocities, and accelerations from other vehicles \cite{Kennedy_platooning}. 
Due to the interconnections among subsystems in distributed models, graph theory \cite{BoChen_Distr_TAC2,YuchenZhang_Distr_TCYB} can be exploited, and the surveys on DSE can be found in \cite{Rego_Distr,Primadianto_Distr,Ge_Distr}. 
Moreover, in the multi-sensor case, the measurement equation is expressed as $y_{i}(t)=C_{i}x_{i}(t)+v_{i}(t)\ (i=1,\cdots,L)$, where $L\in\mathbb{Z}_{+}$ denotes the total number of sensors. 
Multi-sensor fusion methods have been extensively used in target tracking systems \cite{XiaorongLi_Fusion}, and the surveys on multi-sensor fusion estimation can be found in \cite{XiaorongLi_Fusion,Smith_Fusion,Li_Fusion,Sun_Distr2}.

\begin{figure*}[ht]         
    \centering
    \includegraphics[width=18cm]{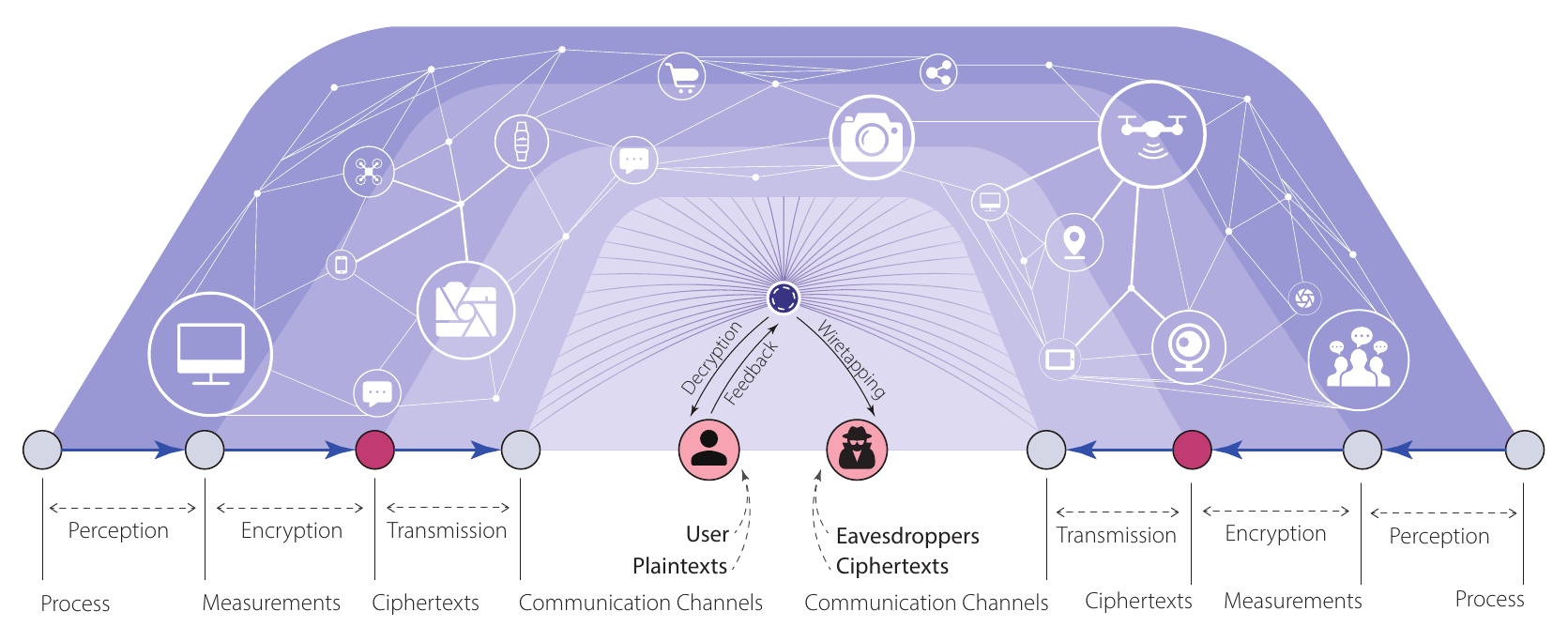}
    \caption{The privacy-preserving state estimation structure in the presence of eavesdroppers. 
    The sensors firstly perceive the process or the system to generate the raw measurements. 
    After encryption at smart sensors, the ciphertexts will be transmitted through communication networks, in which the legitimate user and eavesdropper can capture the data packets. 
    Generally, the legitimate user obtains more information and can decrypt it into plaintexts, while the eavesdropper can only get ciphertexts.}
\label{Fig_Structure}
\end{figure*}

For monitoring the real-time dynamics, it is vital to estimate the system states based on the measured data from sensors. 
An example of state estimation is shown in Fig. \ref{Fig_ITS}, where unmanned aerial vehicles (UAVs), unmanned ground vehicles (UGVs), and aerial vehicle carriers (AVCs) \cite{Huang_Carrier} cooperate to achieve a common objective, such as their own localization and tracking of some targets. 
In order to compute state estimates as precisely as possible, we need to collect all the available data of the considered system. 
The complete information set can be generally expressed by $\mathcal{I}(t)\triangleq\{y(1),\cdots,y(t),o(1),\cdots,o(t)\}$, which consists of all the measurements $y(t)$ and other information $o(t)$ such as acknowledgments (ACKs) of successful reception \cite{Leong_Drop_TAC}. 
Based on this data, the unbiased estimate can be calculated as $\hat{x}(t)\triangleq\mathbb{E}\{x(t)|\mathcal{I}(t)\}$, while the most widely used state estimators in the literature are Kalman-like filters \cite{Kalman}, whose structure is as follows: 
\begin{eqnarray}                              
  \begin{aligned}
  \label{hat_x}
    \hat{x}(t)=A\hat{x}(t-1)+K(t)(y(t)-CA\hat{x}(t-1))
  \end{aligned}, 
\end{eqnarray}
where, at time $t$, $\hat{x}(t)$ stands for the state estimate and $K(t)$ denotes the estimator gain. 

In most cases, the estimator performance index is set as the minimum means square error (MMSE) sense, which means minimizing $P(t)\triangleq\mathbb{E}(\tilde{x}^{\mathrm{T}}(t)\tilde{x}(t)|\mathcal{I}(t))$ with $\tilde{x}(t)\triangleq x(t)-\hat{x}(t)$. 
This estimator is called the Kalman filter (KF) and was proposed in 1960 by Rudolf Kalman \cite{Kalman}. 
Because the KF is recursive, when disturbances affect the system at some given time, their adverse impact will propagate to all subsequent times. 
Hence, by employing the measurement at the current time slot, the state estimation problem can be formulated as a least square (LS) problem. 
The weighted LS (WLS) estimator is one of the most popular forms \cite{Dehghanpour_WLS,Wang_WLS}, which is designed as $\hat{x}(t)=\mathrm{argmin}_{x}(\varepsilon^{\mathrm{T}}(t)\varepsilon(t)|\mathcal{I}(t))$ with $\varepsilon(t)\triangleq y(t)-\hat{y}(t)$. 
Besides, some other indexes or forms are also considered under certain assumptions, such as minimum squared error (SE) \cite{BoChen_Distr_TAC}. 

During the transmission process, numerous malicious eavesdroppers attempt to access the information of the estimation systems. 
The eavesdroppers can employ similar techniques to the legitimate users to demodulate and decode the transmitted packets from communication networks. 
Their ultimate goal is to obtain the real-time accurate states of the systems. 
Nonetheless, in most scenarios, only raw sensed information is sent to the legitimate users and subsequently processed for calculating state estimates. 
Consequently, the transmitted data cannot directly reflect the system state, and eavesdroppers often imitate the operations executed by legitimate users. 
For example, after intercepting raw local measurements, eavesdroppers will apply the KF for the estimate calculation \cite{Leong_Drop_TAC}. 

In reality, we cannot acquire the detailed operations of eavesdroppers since they primarily engage in passive wiretapping. 
Therefore, most papers investigated privacy-preserving approaches against various assumptions about transmission and eavesdropping strategies. 
With cryptography \cite{Rivest_HE}, the sensors can encrypt the data into complex ciphertexts while the legitimate users can decrypt them with secret keys. 
Due to the lack of necessary secret keys, the eavesdroppers cannot successfully decrypt the received information, and what they acquire are garbled codes. 
In some perturbation methods \cite{XinhaoYan_DP_TAES}, sensors inject extra generated noises into originally transmitted values, where eavesdroppers cannot acquire high estimation performance even with KF. 
For scheduling problems \cite{Leong_Drop_TAC}, the local information will be decided whether to be sent or not. 
In this paper, we collectively refer to the privacy-protecting operations in sensors as encryption processes, because these procedures are used to impact the eavesdroppers, and the finally transmitted data are called ciphertexts. 
Moreover, the legitimate users sometimes give feedback to the sensors such as ACKs \cite{Leong_Drop_TAC}. 
Hence, the comprehensive privacy-preserving state estimation structure in the presence of eavesdroppers is demonstrated in Fig. \ref{Fig_Structure}, where the sensor transmitted the ciphertexts to communication channels, in which the users and eavesdroppers acquire information.

\section{Cryptography} 
Cryptographic methods are effective in preserving the accuracy of transmitted messages, because the decryption procedures relying on secret keys are designed to completely recover the original data \cite{Goldreich_Crypt,Goldreich_Crypt2}. 
This kind of method is valuable for maintaining the high performance of legitimate users while guaranteeing sufficient privacy. 
Nevertheless, most of these methods require high computation amounts since the secret keys should be long and complex enough to counter the possible decryption of eavesdroppers equipped with high-performance devices. 
Accordingly, cryptographic methods are commonly considered for the systems that require high legitimate performance and have enough computation resources.

\begin{table*}
\centering
\fontsize{8}{10}\selectfont
\caption{Homomorphic Encryption for Kalman-like estimators.}
\begin{tabular}{|c|c|c|c|c|c|c|c|c|}
\hline
Literature 
& \cite{Gonzalez_HE}
& \cite{Landa_HE}
& \cite{Zhang_HE}  
& \cite{Aristov_HE}
& \cite{Ni_HE}  
& \cite{Emad_HE}  
& \cite{Ristic_HE}  \\
\hline
Estimator 
& EKF 
& KF 
& KF  
& IF 
& KF  
& CFE \& DFE  
& CI  \\
\hline
HE
& additive 
& additive 
& hybrid  
& additive 
& additive  
& additive  
& additive  \\
\hline
\end{tabular}
\label{Table_HE}
\end{table*}

\subsection{Homomorphic Encryption} 
Homomorphic encryption (HE) \cite{Rivest_HE} is a classical cryptographic approach that has been utilized for many applications \cite{Geva_HE}. 
It allows certain operations on the encrypted data, and more concretely, the operations on ciphertexts can be reflected on the plaintexts. 
In general, there are two homomorphisms: additive homomorphism and multiplicative homomorphism, which respectively means addition and multiplication are operable on ciphertexts, and the detailed explanations are given as follows. 

\begin{enumerate}
\item \emph{Additive Homomorphism}: 
Certain operations on ciphertexts can calculate the addition on plaintexts, i.e.,  
\begin{eqnarray}                                           
  \begin{aligned} 
  \label{Add}
    \mathcal{D}(\mathcal{E}(m_{1},\mathrm{pk})\oplus\mathcal{E}(m_{2},\mathrm{pk}),\mathrm{sk})=m_{1}+m_{2}
  \end{aligned}, 
\end{eqnarray}
Here, $m_{1}$ and $m_{2}$ are two plaintexts. 
$\mathcal{E}(\cdot)$ and $\mathcal{D}(\cdot)$ respectively stand for the encryption procedure and decryption procedure, while ``$\mathrm{pk}$'' and ``$\mathrm{sk}$'' are the respective public key and private key. 
The symbol $\oplus$ means the operation reflects the addition on plaintexts. 
\item \emph{Multiplicative Homomorphism}: 
Certain operations on ciphertexts can calculate the multiplication on plaintexts, i.e., 
\begin{eqnarray}                                           
  \begin{aligned} 
  \label{Mlt}
    \mathcal{D}(\mathcal{E}(m_{1},\mathrm{pk}) \otimes \mathcal{E}(m_{2},\mathrm{pk}),\mathrm{sk})=m_{1}m_{2}
  \end{aligned}, 
\end{eqnarray}
where the symbol $\otimes$ stands for the operation related to the multiplication on plaintexts. 
\end{enumerate}

Depending on whether both the two homomorphisms are satisfied, homomorphic schemes can be divided into two main kinds: fully homomorphic encryption (FHE) \cite{Gentry} and partially homomorphic encryption (PHE) \cite{Paillier,RSA,ElGamal}. 
FHE \cite{Gentry} can simultaneously support both additive and multiplicative homomorphisms, while PHE only supports one of them. 
Paillier \cite{Paillier} is a kind of classical additive HE, while RSA \cite{RSA} and ElGamal \cite{ElGamal} are typically multiplicative HE approaches. 
Besides, there also exists a special type called hybrid HE \cite{Zhang_HE}, where two different HE approaches are combined to respectively achieve multiplicative and additive homomorphisms. 
Note that FHE requires a long computation time, which has been demonstrated by the practical experiments in \cite{Abdelhafez_Compare}. 
In those experiments, the estimation with PHE only requires several milliseconds, while that with FHE takes several hours, and that with garbled circuits takes several minutes. 
This implies that PHE is more practical for real-time state estimation in the majority of cases. 

The idea of HE has been widely adopted for various scenarios, including cloud-based systems \cite{Darup_HE_Cloud,Alexandru_HE_Opt,Alexandru_HE_LQG}, Internet of Things (IoT) \cite{Hijazi_HE_IoT}, and intelligent transportation systems (ITSs) \cite{Sultan_HE_Image,Zhu_HE_ITS}. 
Alexandru et al. introduced PHE into cloud-based systems, where a strong cloud server was deployed for secure computation with the information gathered from some weak agents. 
Then, they solved the private cloud-based quadratic optimization \cite{Alexandru_HE_Opt} and linear quadratic Gaussian (LQG) control \cite{Alexandru_HE_LQG} problems. 
Hijazi et al. \cite{Hijazi_HE_IoT} applied FHE to develop a private learning method, where a global model was trained with encrypted model updates from distributed IoT devices. 
Sultan et al. \cite{Sultan_HE_Image} employed Paillier HE to encrypt the searching algorithm over the image gathered from the camera deployed in an autonomous vehicle, and Zhu et al. \cite{Zhu_HE_ITS} encrypted the learning approach with the sensed data from edge devices and road side units. 
Besides, there are plenty of other HE applications, which can be found in the recent surveys \cite{Marcolla_FHE,Acar_HE}. 

In the state estimation field, Gonzalez-Serrano et al. \cite{Gonzalez_HE} applied the additive HE to an extended Kalman filter (EKF) by encrypting the transmitted measurements, and they also developed secure division and matrix inversion protocols. 
Landa and Akbar \cite{Landa_HE} proposed a secure KF for a linear time-invariant system with additive homomorphism. 
Since they did not encrypt system matrices, the above-mentioned secure division or matrix inversion protocols were not required. 
Recently, Zhang et al. \cite{Zhang_HE} also studied a secure steady-state KF, but they utilized hybrid HE for simultaneously protecting the model parameters, measurements, and estimates. 
The raw measurements were firstly encrypted by multiplicative HE and then directly operated by the estimator. 
After decryption with private key of multiplicative HE, the user further encrypted the results with additive HE, and finally obtained the state estimate by resorting to the additions in estimator. 

For the multi-sensor scenario, Aristov et al. \cite{Aristov_HE} exploited the Paillier HE to a multi-sensor information filter (IF). 
Such a secure IF was more efficient than standard HE-based KF thanks to its additive structure about measurement update equations. 
Ni et al. \cite{Ni_HE} considered a situation when an average estimate was calculated by combining all the LSEs in the cloud server, and such an average was then broadcasted and synchronized at each sensor. 
The additive HE was directly embedded, because the average fusion form is a typical kind of summation. 
Meanwhile, they introduced model privacy, a concept in machine learning, to state estimation field for the first time. 
Moreover, Emad et al. \cite{Emad_HE} considered both privacy-preserving centralized and distributed fusion estimators against sensor, cloud, and query coalitions. 
The Paillier HE was exploited to encrypt estimates and measurements, and the multi-party privacy with respect to semi-honest parties was further proved by resorting to the definitions about computational indistinguishability in cryptography.
In addition, the HE-based covariance intersection (CI) fusion structure was discussed by Ristic \cite{Ristic_HE}, and an approximate solution to fast CI was derived by resorting to order revealing encryption. 
Besides, we summarize the HE study for Kalman-like estimators in Table. \ref{Table_HE}. 

Except for KF and its extended versions, HE has also been studied for other state estimators. 
Mohsen et al. \cite{Mohsen_HE} applied Paillier HE to the Luenberger observer (LO) and further derived the implementation condition on practical devices with limited memory sizes. 
Then, a distributed LO under a canonical decomposition was encrypted by Kim et al. \cite{Kim_HE} with additive HE. 
Alanwar et al. \cite{Alanwar_HE} proposed a secure set-based state estimator when the sets are represented by zonotopes. 
They also studied centralized and distributed fusion scenarios with distributed sensors and sensor groups, respectively. 

Furthermore, it is crucial to note that the data intended for processing must reside within the plaintext space. 
This prerequisite is vital for the successful implementation of HE, but it is often overlooked. 
For example, the estimation performance degradation was neglected for other theoretical analyses in \cite{Ni_HE}. 
In order to convert data into plaintexts, the floating-point number was expressed by a positive exponent and a mantissa in \cite{Ziad_HE}, and the resultant representations were all integers that were able to be encrypted. 
This encoding method has already been employed to the state estimation field \cite{Emad_HE,Alanwar_HE}. 
Instead of the detailed representation of the whole float number, quantization $Q:\mathbb{R}\to\mathbb{Z}$ is another method to transfer the data into the space of plaintext \cite{Zhang_HE,Kim_HE}. 
In general, the quantizer only preserves the integer part and directly deletes the mantissa. 
For instance, a probabilistic uniform quantization was adopted in \cite{Zhang_HE}, where the output was chosen randomly in an interval. 

\subsection{Other Cryptographic Methods} 
Secure multi-party computing (SMPC) is a branch of cryptography focusing on enabling multiple parties to collectively compute desired functions while guaranteeing the privacy of their own inputs.
It allows the parties to jointly perform computations on their private data without revealing any sensitive information to each other. 
The concept of SMPC was proposed by Yao in \cite{Yao_SMPC}, where the millionaire’s problem was discussed. 
HE is a typically effective method to achieve SMPC because of the homomorphisms, and various HE-based estimators have been proved to satisfy SMPC  \cite{Alexandru_HE_Opt,Alexandru_HE_LQG,Emad_HE,Alanwar_HE}. 
More concretely, multi-party privacy with respect to semi-honest behavior can be realized by analyzing the computationally indistinguishablity of excution views \cite{Alanwar_HE}. 
The survey on SMPC can be found in \cite{Cramer_SMPC}. 
Except for HE, secret sharing (SS) is another useful cryptographic approach \cite{Pedersen_SS}. 
A typical scheme is additive 2-out-of-2 SS, where the message is split into two parts \cite{Hofmeister_SS}. 

Another classical cryptographic technique is watermarking that can transfer the received information into original data without affecting the original accuracy \cite{Iacovazzi_Water,Song_Water}.
A pseudo-random generator is a deterministic function that can expand short seeds into longer sequences of pseudo-random bits, and the encryption with pseudo-random sequence is a kind of effective watermarking method \cite{Niederreiter_Pseudo}. 
Ristic et al. \cite{Ristic_Pseudo} exploited a stream of pseudo-random Gaussian samples to degrade the performance at unprivileged estimators, while maintaining the performance at the privileged estimation. 
Huang et al. \cite{Huang_Water} added the pseudo-random sequence to the constantly linear transformation of innovation so that the $\chi^{2}$ detector can successfully detect linear man-in-the-middle attacks. 
Furthermore, the watermarking with pseudo-random samples was considered for distributed state estimation by Zhou et al. \cite{Zhou_Water}, where the interactive neighbouring estimate was multiplied by a watermarking matrix composed of some pseudo-random numbers. 
Besides, the Kullback–Leibler (K–L) divergence detector in \cite{Zhou_Water} was demonstrated more powerful resistance against stealthy attacks when compared to that in \cite{Huang_Water}.

\subsection{Discussion} 
With the assistance of secret keys, key owners can consistently decrypt the ciphertexts into their corresponding plaintexts. 
Particularly, based on certain homomorphisms, it is possible to do some operations on the complex ciphertexts, thereby enhancing the flexibility of cryptographic approaches. 
Therefore, by resorting to the tools of cryptography, legitimate users can obtain the desired data with relatively high accuracy. 
Meanwhile, the data privacy can be strongly guaranteed on account of the complexity theory, and the eavesdropper cannot acquire any valid information without secret keys. 

Nevertheless, some cryptographic approaches still cause minor errors during preparation, encryption or decryption. 
Specifically, homomorphic encryption can only
operate on integers, because its encryption and decryption functions are based on modular arithmetic,
which is a type of arithmetic that deals with integers \cite{Alanwar_HE}. 
Also, the secure transmission and preservation of secret keys are important problems, because the secret keys should be kept in fully trusted devices. 
For instance, the distribution of the seed of pseudo-random sequence should be entirely secure, otherwise, it is possible for the eavesdropper to completely recover the original information. 

Moreover, most cryptographic methods require large computations for successful realization due to the large key sizes and complex exponential operations. 
In many practical systems, such a large computation burden will affect their normal operations, especially real-time processes such as estimation and control. 
For example, when the authority period of a UAV is set as 1 ms while a cryptographic method needs 10 ms at each time slot, the encryption procedures cannot be completely executed and even occupy the time for control. 
Particularly, the experiments have demonstrated that FHE requires much more time than PHE \cite{Abdelhafez_Compare}, and that is why most of the existing estimators considered PHE. 
Thus, it is of great demand to derive more computationally efficient cryptographic approaches.

\section{Data Perturbation}
Data perturbation is an extensively utilized privacy-preserving method mainly related to increasing the uncertainty of specific information. 
In a sense, it bears a resemblance to cryptography, as both strive to alter the values of data to confuse potential eavesdroppers. 
By contrast, perturbation techniques are always simpler, enabling the eavesdroppers to consistently access the obfuscated data. 
In other words, even though the eavesdroppers can successfully decode the packages, their finally obtained data are also inaccurate. 
A great benefit of data perturbation is that the computation consumption is minor since only some simple data modifications instead of long strings of secret keys are required. 
For disturbing information, random noise injection is one of the most effective methods. 
Notice that it is different from the pseudorandom sequence in cryptography \cite{Ristic_Pseudo}, because the true random samples are unknown to the legitimate user instead of known as secret keys. 
In this case, the disturbance may simultaneously degrade the performances of legitimate users. 
Hence, it is required to pay attention to the utility of data while guaranteeing privacy with perturbation methods.

\begin{table*}
\centering
\fontsize{8}{10}\selectfont
\caption{Differential Privacy for Kalman-like estimators.}
\begin{tabular}{|c|c|c|c|c|c|c|c|c|c|c|c|c|}
\hline
Literature 
& \cite{Ny_DP_TAC}
& \cite{Yazdani_DP,Degue_DP}
& \cite{XinhaoYan_DP_TAES,XinhaoYan_DP_Auto}  
& \cite{Yan_DP}
& \cite{Dawoud_DP}  
& \cite{Wang_DP_UKF}  
& \cite{Andre_DP_EnKF}  
& \cite{Vishnoi_DP_MHE}\\
\hline
Estimator 
& KF 
& KF 
& DFE  
& CFE 
& set-based  
& UKF  
& EnKF 
& MHE  \\
\hline
Mechanism
& Laplace \& Gaussian 
& Gaussian 
& Gaussian  
& Gaussian
& truncated Laplace
& Laplace
& Gaussian
& Gaussian  \\
\hline
\end{tabular}
\label{Table_DP}
\end{table*}

\subsection{Differential Privacy} 
Differential privacy (DP) is known as a classical perturbation method and was proposed by Dwork et al.\ in 2006 \cite{Dwork_DP,Dwork_DP2}. 
The artificial noises with certain distributions are required to be injected for disturbing the statistical information of a great number of elements in a database. 
Define a space $D$ satisfying a adjacent relation $\mathrm{Adj}(d,d')$, which means that $d\in D$ and $d'\in D$ only differ by certain components. 
Let $(R,\mathcal{M})$ be a measurable space, and a mechanism $M$ is differentially private for all $d$ and $d'$ such that $\mathrm{Adj}(d,d')$, one has \cite{Ny_DP_TAC} 
\begin{eqnarray}                                  
  \begin{aligned} 
  \label{DP}
    \mathbb{P}(M(d)\in S)\leq e^{\varepsilon}\mathbb{P}(M(d')\in S)+\delta,\ \forall S\in\mathcal{M}. 
  \end{aligned}
\end{eqnarray}
In other words, this performance indicates that the distributions over the outputs of the mechanism are close to some extent for two adjacent datasets. 
The relevant works have been studied a lot, and the surveys can be found in \cite{Hassan_DP}. 
Here, we simply introduce recent theoretical developments and then elaborate on the research about differentially private estimators. 

To achieve the above index, plenty of noise mechanisms have been discussed, such as Gaussian \cite{Dwork_DP2014}, Laplace \cite{Dwork_DP}, uniform \cite{Geng_DP}, and exponential \cite{McSherry_DP}. 
These mechanisms concentrate on the random noise addition, i.e, $M(d)=q(d)+n$, which means injecting noise $n$ into the original query output $q(d)$. 
Here, we briefly address the classical Gaussian and Laplace mechanisms in regular use. 
If a random variable satisfies Gaussian distribution  $x\sim\mathcal{N}(0,\sigma^{2})$, its probability density function (PDF) is $p(x;\sigma)=\frac{1}{\sqrt{2\pi}\sigma}e^{-\frac{x^{2}}{2\sigma^{2}}}$. 
Then, a white Gaussian noise  $n\sim\mathcal{N}(0,\sigma^{2}I)$ satisfying $(\varepsilon,\delta)$-DP when the following condition holds: 
\begin{eqnarray}                                  
  \begin{aligned} 
  \label{Gaussian}
    \sigma\geq\frac{\Delta_{2,q}}{2\varepsilon}(K+\sqrt{K^{2}+2\varepsilon}),
  \end{aligned}
\end{eqnarray}
where $K=Q^{-1}(\delta)$ with $Q(x)\triangleq\frac{1}{\sqrt{2\pi}}\int_{x}^{\infty}e^{-\frac{u^{2}}{2}}du$ and $\Delta_{2,q}\triangleq\mathrm{sup}_{\mathrm{Adj(d,d')}}\|q(d)-q(d')\|_{2}$ is the sensitivity. 
Similarly, a Laplace variable $x\sim\mathcal{L}(0,b)$ has the PDF $p(x;b)=\frac{1}{2b}e^{-\frac{|x|}{b}}$. 
If a Laplace noise $n\sim\mathcal{L}(0,b^{k})$ satisfies $\varepsilon$-DP, one has 
\begin{eqnarray}                                  
  \begin{aligned} 
  \label{Laplace}
    b\geq\frac{\Delta_{1,q}}{\epsilon}.
  \end{aligned}
\end{eqnarray}

Also, Sadeghi et al. \cite{Sadeghi_DP} proposed an offset symmetric Gaussian tails mechanism, and the detailed distribution was acquired by using the normalized tails of two symmetric Gaussian distributions. 
Alghamdi et al. \cite{Alghamdi_DP} proposed cactus mechanisms, where the optimization target that minimizes the Kullback-Leibler divergence was quantized for calculating near-optimal mechanisms. 
Muthukrishnan et al. \cite{Muthukrishnan_DP} introduced a hybrid addition mechanism that resembled a Laplace distribution in the center and a Gaussian distribution in the tail. 
The necessary and sufficient conditions for one dimension as well as a sufficient condition for certain dimensions were further derived. 
Then, Kadam et al. \cite{Kadam_DP} applied a tractable structure to design the probability mass function (PMF) of noise, which optimized the expected distortion with certain differentially private parameters. 
Balle et al. \cite{Balle_DP} argued that the variance design of the previous mechanism was not precise enough for the scenario demanding high privacy level. 
They proposed an optimal Gaussian mechanism calibrating the variance with Gaussian cumulative density function (CDF) instead of approximating tail bounds. 

Thanks to the powerful performance and rigorous mathematical models, DP has been expanded to a wide range of fields, such as machine learning \cite{Han_DP_Learning}, optimization \cite{Han_DP_Optimization}, consensus \cite{Liu_DP_Consensus}, game theory \cite{Ye_DP_Nash} and control theory \cite{Kawano_DP_Control,Wang_DP_Control,Degue_DP_Control}. 
Concretely, Han et al. \cite{Han_DP_Learning} proposed a differentially private distributed dual averaging method for the unbalanced directed networks.
The privacy index was dependent on the variance of injected Laplace noises and the Lipschitz constant of the proposed loss functions. 
Liu et al. \cite{Liu_DP_Consensus} proposed a differentially private consensus control approach for the heterogeneous multi-agent systems. 
A time-varying control gain was designed to mitigate the decaying effect, and the non-decaying noises were injected to avoid the privacy leakage of reference state. 
Wang et al. \cite{Wang_DP_Control} showed that the performance cost of this mechanism was related to the time horizon and the privacy parameter for stable systems, while the cost increased with time exponentially for unstable systems. 
Kawano et al. \cite{Kawano_DP_Control} demonstrated the relationship between DP and the input observability or left invertibility of a system. 
More specifically, the Gaussian mechanism effectively enhanced DP by injecting relatively small disturbance when the system was less input observable. 
As a result, they proposed a confidential dynamic controller by adding minor noise, achieving superior privacy level. 
For a multi-agent system, Degue and Ny \cite{Degue_DP_Control} proposed a LQG controller based on measurement aggregation, where a KF was adopted to reduce the negative impact of noise mechanism on aggregation. 

Particularly, in estimation and filtering theory, the differentially private estimator was firstly proposed by J. Le Ny \cite{Ny_DP,Ny_DP2,Ny_DP_TAC,Ny_DP_Book}. 
According to the fundamental definitions in database field, they gave the adjacency relation on states in system theory: 
\begin{eqnarray}                                 
  \begin{aligned} 
  \label{Adj}
    &\mathrm{Adj}(x,x')\ \mathrm{iff\ for\ some}\ i,\ \|S_{i}x_{i}-S_{i}x_{i}'\|\leq\rho_{i}  \\
    &(I-S_{i})x_{i}=(I-S_{i})x_{i}',\ \mathrm{and}\ x_{j}=x_{j}'\ \forall j\neq i 
  \end{aligned}, 
\end{eqnarray}
where $S_{i}$ is the selection matrix with binary diagonal components. 
It means that the distance between the adjacent state acquired by the eavesdroppers and the real state of the system is bounded. 
Based on this assumption, the sensitivity in mechanism \eqref{Gaussian} or \eqref{Laplace} can be easily obtained, thus deriving the noise variance. 

\begin{figure}[ht]         
    \centering
    \includegraphics[width=8cm]{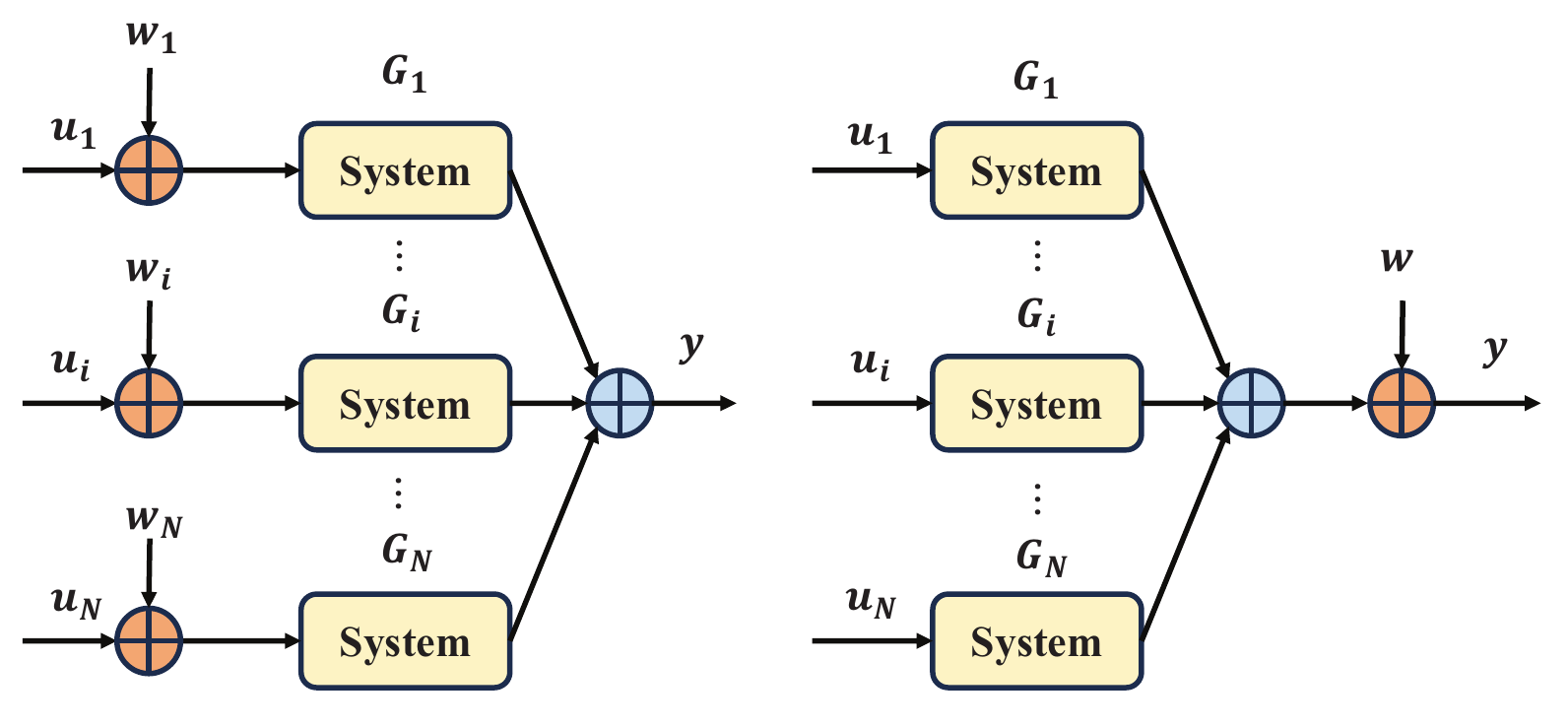}
    \caption{Two perturbation structures of DP in system theory.}
\label{Fig_Perturbation}
\end{figure}

Notice that the final aim is to perturb the statistical information of query outputs. 
Hence, instead of directly adding noises to the system output, the noises can be inserted into the system inputs \cite{Ny_DP_TAC}. 
The two perturbation structures in system theory are demonstrated in Fig. \ref{Fig_Perturbation}. 
Moreover, by expanding this definition to the multi-sensor systems, these two structures respectively stand for the noise injection on local sensors and FC \cite{XinhaoYan_DP_TAES}.

Building on the above fundamental architecture \cite{Ny_DP_TAC}, numerous extensive studies about differentially private estimators have been conducted. 
For the input perturbation scheme, Yazdani and Hale \cite{Yazdani_DP} proposed the bounds of the prior and posterior estimation errors, as well as that of the differential entropy. 
Similarly, Degue and Ny \cite{Degue_DP_Conf,Degue_DP} proposed a two-stage structure that aggregates distributed information before inserting random noises. 
The estimator parameters were optimized by minimizing the mean square error (MSE) and the optimization problem was solved by a semi-definite program. 
Then, the multi-sensor case was studied by Yan et al. \cite{XinhaoYan_DP_TAES}, where each sensor had the possibility of injecting local noises. 
A gain related to the weighting fusion matrix was applied to transfer the mechanism into the form of white noise addition, enabling that the noises can be directly designed with the existing mechanisms in \cite{Ny_DP_TAC}. 
Then, Yan et al. further applied noises to simultaneously preserve the privacy of LSEs and DFE \cite{XinhaoYan_DP_Auto}. 
Inspired by the moving average (MA), the summation form about LSEs was proposed for being transmitted in the unreliable communication networks. 
Two-step sequential noises were injected into the summation results, and both of them were combined for realizing DP. 
Besides, to reduce the adverse effect of unknown noises, an extra privatized sensor was deployed in \cite{Yan_DP} to ensure the fusion estimation performance. 

In addition to the standard KF, a substantial amount of other estimators were discussed along with DP, including multi-input multi-output (MIMO) filter \cite{Ny_DP_MIMO}, interval observer (IO) \cite{Degue_DP_Interval}, and many nonlinear estimators \cite{Wang_DP_UKF,Andre_DP_EnKF,Ny_DP_Nonlinear,Vishnoi_DP_MHE}. 
In \cite{Ny_DP_MIMO}, Le Ny and Mohammady proposed several approximation forms of MIMO filters for guaranteeing DP. 
They also derived an optimal zero-forcing equalization (ZFE) mechanism to approximate single-input multiple-output (SIMO) filters and a suboptimal diagonal prefilter for general MIMO cases. 
Degue and Le Ny \cite{Degue_DP_Interval} proposed a confidential IO for a multi-agent system under generally bounded disturbances, and the truncated Laplace noises were added to the transmitted data at each agent. 
Similar to the above work, Dawoud et al. \cite{Dawoud_DP} considered a privacy-preserving set-based estimator with zonotopes for set representation, where DP was achieved with the above truncated Laplace mechanism. 

Moreover, Wang et al. \cite{Wang_DP_UKF} proposed a differentially private unscented KF (UKF) to protect the privacy of user in IoT. 
After aggregating streaming data and injecting Laplace noises, the UKF was adopted to reduce the noise influence. 
Andre and Ny \cite{Andre_DP_EnKF} applied a Gaussian mechanism to ensure the confidentiality of individual drivers, where the traffic density or velocity should be estimated. 
In a similar way, the noises were inserted into the measurements from static detector, and an ensemble KF (EnKF) was employed to increase the estimation accuracy. 
Besides, Ny \cite{Ny_DP_Nonlinear} applied DP to a nonlinear Luenberger-type observer, where the stability was guaranteed and the level of noise was set with contraction analysis. 
Vishnoi et al. \cite{Vishnoi_DP_MHE} exploited the Gaussian mechanism to a moving horizon estimator (MHE). 
The traffic density and relative flow of each road segment were estimated, and the noise mechanism was derived on the basis of the sensitivities for measured states. 
Among these nonlinear methods, the estimation performance of UKF was the best under certain privacy parameters, followed by MHE and EKF, while that of EnKF was the worst \cite{Vishnoi_DP_MHE}. 
The DP study for Kalman-like estimators is shown in Table. \ref{Table_DP}. 

According to the existing literature, the integration of differentially private mechanisms into system theory presents a fatal drawback, that is the legitimate user can just utilize the disturbed data as what eavesdroppers receive. 
Unlike the process of decrypting data with cryptography, the disturbances caused by the integration of DP cannot be entirely eliminated. 
Such perturbation will lead to the degradation in estimation accuracy for legitimate estimators, which is a crucial performance metric for estimation systems. 
Generally, the negative impact of such uncertainty can be decreased with estimation methods such as KF \cite{Ny_DP_TAC} and weighting fusion \cite{XinhaoYan_DP_TAES}. 
In a specific case, the noises can be designed with the null space of the weighting matrices such that the noise sequences were discarded \cite{XinhaoYan_DP_Auto}. 

There was also recent work by Murguia et al. \cite{Carlos2} where the data \textit{and the estimator run by the user} were transformed using random affine transformations before disclosure. 
The transformed estimator was designed to work on the distorted data to produce distorted state estimates, which were fed back to the user. 
Then, the user extracted the correct estimate from the distorted one using the inverse of the transformation (a scheme similar to homomorphic encryption). 
Furthermore, this scheme was proved to provide an arbitrary level of DP without distorting the state estimates.

\subsection{Other Perturbation Methods} 
In addition to meeting the index requirements of DP, the direct application of random noises also proves to be effective. 
Most generally, the artificial noises were directly injected into the measurements by Das and Bhattacharya \cite{Das_Noise}, and the privacy-utility tradeoff was solved by formulating optimization problems. 
Chen et al. \cite{Chen_Noise} proposed a stealthy perturbation strategy that prevented the eavesdropper from detecting the exception and acquiring precise data. 
The random noises were injected into the local measurements, but the variation of innovation was still bounded after such perturbation. 
Subsequently, the limited number of encrypted sensors was calculated by solving an optimization problem. 
Moradi et al. \cite{Moradi_DP_Distr} proposed a privacy-preserving distributed KF for a multi-agent system by resorting to the techniques about average consensus. 
The considered states were divided into private and public parts, while only the public ones were perturbed by the Gaussian noises. 

Moreover, Yan et al.\cite{XinhaoYan_DP_Auto} injected two-step sequential noises the moving averages about LSEs such that the difference-based estimation errors of the eavesdropper grew unbounded. 
Meanwhile, the partial noise was eliminated with the null space design and another part was compensated in the linear minimum variance sense, ensuring the stability of the fusion estimator for the legitimate user. 
Similarly, Yan et al. \cite{XinhaoYan_Eaves_TAES} proposed a white noise injection approach by resorting to the null space related to the fusion criterion in DFE. 
Different from \cite{XinhaoYan_DP_Auto}, the noises designed in \cite{XinhaoYan_Eaves_TAES} were employed to damage the fusion estimator of the eavesdropper, because the noises were related to the eavesdropper's weighting fusion matrix. 
This means that the eavesdropper lost the information of certain components under such encryption scheme, resulting in a degradation of its fusion estimation performance. 

Apart from direct random perturbation, there are other modifications to the transmitted data that can also safeguard the confidentiality of the systems. 
A typical method is called compressive privacy introduced by Kung et al. \cite{Kung_Compressive}, applying a linear transformation or compression for mapping the desired data into a lower-dimensional space.
Song et al. \cite{Song_Compressive} devised an optimization problem that balances the current system utility with privacy considerations for multiple future steps. 
Then, they illustrated the method to determine the optimal compressive matrix, while in the multi-sensor scenario, each sensor independently optimized its local compressive mapping. 
In fact, the null space idea in \cite{XinhaoYan_Eaves_TAES} is similar to compressive privacy, because certain components are deleted under the proposed mechanism. 

Shang et al. \cite{Shang_Linear} employed linear encryption to defend against the stealthy attacks, which used linear transformation to avoid being detected by the $\chi^{2}$ test approach \cite{Guo_Linear}, and the measurements or innovations were operated with a scaling matrix and a bias vector.
Also, Shang and Chen utilized the linear encryption strategy for countering the eavesdropping attacks \cite{Shang_Linear2}, where the estimation error of legitimate use was minimized and the lower bound of the eavesdropper’s steady-state estimation error covariance was maximized.
The covariance of the eavesdropper was proved to grow unbounded for marginally stable or unstable systems in the single-dimensional case.

The tools of information theory can more effectively demonstrate the uncertainty or perturbation, and the related methods have been widely discussed to enhance the privacy level \cite{Nekouei_Review}. 
Some information-theoretical metrics have been considered to measure the target, such as conditional entropy \cite{Nekouei_Information,Nekouei_Information2}, direct information \cite{Tanaka_Information}, and mutual information \cite{Liu_Information,Nekouei_Information3,Carlos1}. 
The discrete conditional entropy was employed to denote the uncertainty by Nekouei \cite{Nekouei_Information}, and an optimal privacy-aware estimator was developed through minimizing the average loss while adhering to a lower privacy-related bound.
Tanaka et al. \cite{Tanaka_Information} derived the optimal joint control and estimation approach with Kramer’s causally conditioned directed information. 
Liu et al. \cite{Liu_Information} minimized the estimation error covariance when the mutual information is subject a boundary for guarantee the sufficient privacy. 
Hayati et al. \cite{Carlos3,Carlos4} formulated the design of privacy-preserving mechanisms as a privacy-utility tradeoff problem, where the synthesis of related Gaussian mechanisms was posed as the solution of a series of convex programs.
They minimized the finite horizon \cite{Carlos3} and infinite horizon \cite{Carlos4} mutual information between the sequence of states and the estimates produced by worst-case eavesdroppers using KFs.

\subsection{Discussion} 
Obviously, the introduction of random factors can reduce the accuracy of processed data, thereby impairing the eavesdropper's estimation performance. 
Different from the complex procedures about key generation and encryption-decryption, perturbation methods can be directly implemented by generating and inserting random numbers. 
Such a straightforward procedure of numerical addition demands fewer computational resources, thus speeding up encryption and saving time for other operations. 

Unlike cryptography, most perturbation methods will simultaneously impact the performance of legitimate users due to the lack of accurate real-time disturbances. 
Therefore, the trade-off discussion becomes essential, which involves the considerations of legitimate users' performance, eavesdroppers' performance, and even energy consumption. 
Nevertheless, it is generally assumed that the unauthorized eavesdroppers and legitimate users have the same knowledge, which is not appropriate enough for estimation systems. 
In some cases, legitimate systems have certain private information that can be applied to enhance the privacy level, such as the initial values set in advance. 
Hence, it is also of great significance to strive to make full use of these unique data to compensate for loss of performance \cite{XinhaoYan_Eaves_TAES} or eliminating the adverse impact \cite{XinhaoYan_DP_Auto}. 

In fact, the eavesdropper can still acquire the rough system state on certain level thanks to the remaining information of the perturbed data \cite{XinhaoYan_DP_TAES}. 
This means that although the data are encrypted with stochastic samples, the low randomness may not completely affect the eavesdropper's performance by considering the data utility for the legitimate user. 
Besides, the relatively low accurate data may be enough for the eavesdropper to launch malicious practical attacks, such as missile attack, but the possible dreadful consequences should not exist. 
Furthermore, the eavesdroppers can even speculate a more precise range about the system state by exploiting other information. 
To summarize, the level of privacy achieved through data perturbation methods is generally lower than that achieved through cryptography, and thus finding ways to enhance it presents a significant challenge.

\section{Transmission Scheduling}
Transmission scheduling is another type of privacy protection method relying on the fundamental assumption that all the communication channels, whether they are constructed by legitimate users or potential eavesdroppers, inherently possess the possibility for packet dropping. 
In most case, the related methods set and optimize the expected targets to enable the design of effective scheduling methods. 
These methods are customized to function within certain predefined constraints, ensuring a balance between utility and privacy indexes. 
Furthermore, it can be beneficial to introduce artificial factors to manipulate the communication channels or scheduling strategy. 
Such operations can significantly increase the data complexity and ambiguity for unauthorized parties, thereby enhancing the privacy level.

\subsection{Random Packet Drop} 
\begin{figure}[ht]         
    \centering
    \includegraphics[width=8cm]{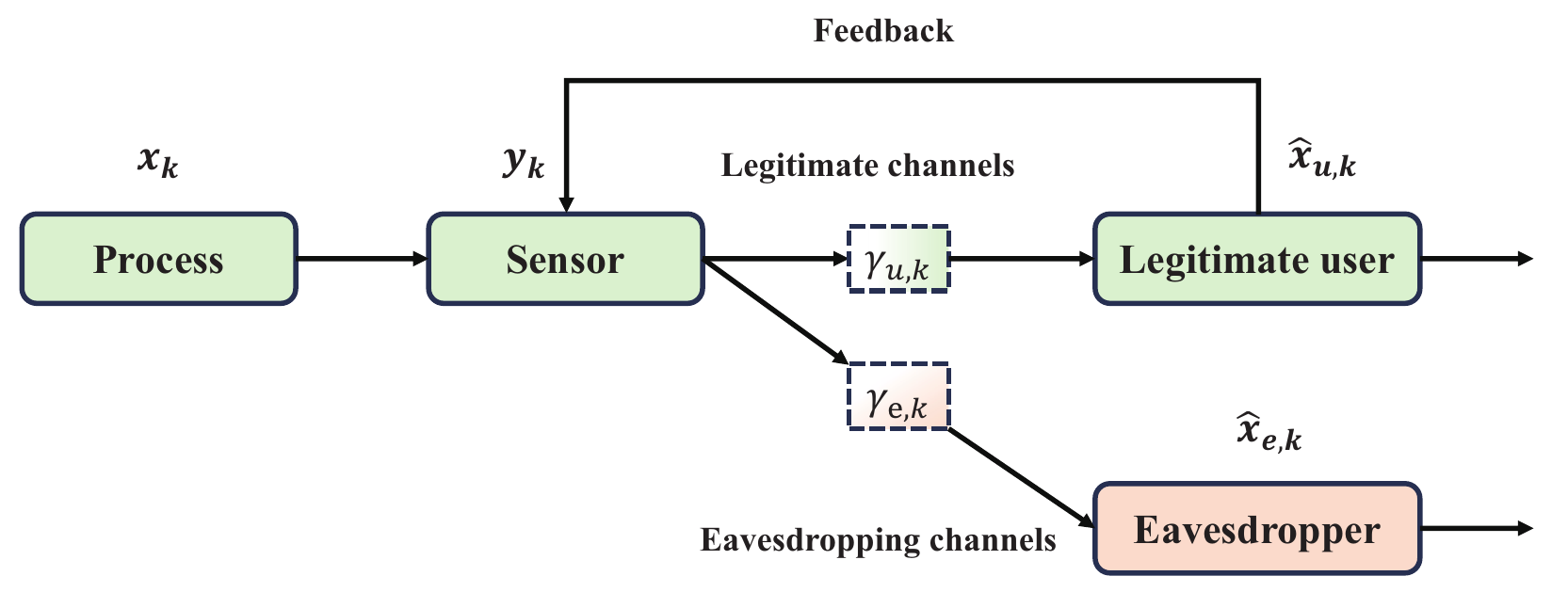}
    \caption{The packet drop structure for estimation systems in the presence of eavesdroppers.}
\label{Fig_Scheduling}
\end{figure}

In the context of this approach, packet drop is an inevitable phenomenon in the networked transmission due to some inherent characteristics of communication systems such as buffer overflow and routing issues. 
Particularly, there is a probability that the local information transmitted between the sensor and the remote estimator may experience dropout \cite{Leong_Drop_TAC}.
Let $\gamma_{u}(t)$ be the random variable such that $\gamma_{u}(t)=1$ when the transmitted package is successfully received by the remote estimator at time $t$, and otherwise $\gamma_{u}(t)=0$.
The subscript ``$u$'' represents the parameters of legitimate users. 
Similar definitions are given for $\gamma_{e}(t)$, where the subscript ``$e$'' stands for the parameters of eavesdroppers. 
The variables $\gamma_{u}(t)$ and $\gamma_{e}(t)$ are respectively assumed to be independently and identically distributed (i.i.d.) Bernoulli such that \cite{Schenato_Drop}: 
\begin{eqnarray}                            
  \begin{aligned}
  \label{drop}
    \mathbb{P}(\gamma_{u}(t)=1)=\lambda_{u},\ \mathbb{P}(\gamma_{e}(t)=1)=\lambda_{e}
  \end{aligned}. 
\end{eqnarray}
The above packet drop structure for estimation systems in the presence of eavesdroppers is demonstrated in Fig. \ref{Fig_Scheduling}.

Notice that under the above assumption, the estimation performances of legitimate users and eavesdroppers will both degrade.
Hence, the optimization target should be designed to consider the performance evaluation of both these parties.
Specifically, the optimization objective can be determined through minimizing the average mean squared error (MSE) of legitimate users, while simultaneously ensuring that the MSE of eavesdroppers remains above a certain threshold. 
This delicate balance can be achieved by employing approached, including stochastic bit flipping \cite{Aysal_Drop} and transmit filter design \cite{Reboredo_Drop}.

Yuan et al. \cite{Yuan_Drop} firstly applied control feedback to enhance security, ingeniously using a malicious jammer to interfere with a malicious eavesdropper, essentially pitting attackers against each other. 
Considering the constraints of limited power and data utility, they framed the problem of energy distribution as an optimization problem.
The Stackelberg game was employed to model it to effectively manage power allocation while ensuring optimal security against potential eavesdroppers.
Then, Leong et al. \cite{Leong_Drop_TAC} proposed the structural results on optimal transmission scheduling, revealing a threshold behavior in the estimation error covariance.
This thresholding behavior indicates that the transmission scheduling strategy changes when the estimation error covariance crosses a certain threshold. 
In addition, Leong et al. \cite{Leong_Drop_LCSYS} also applied mutual information as a measure to depict the performance, and then determined the upper and lower bounds on the revealed data.
Extensions to the basic framework were explored in the literature, including alternative measures of security \cite{Leong_Drop_arXiv} and Markovian packet drops \cite{Leong_Drop_CDC}. 

Moreover, Tsiamis et al. \cite{Tsiamis_Drop_IFAC} considered a random transmission strategy with process noise and channel randomness. 
When the packet reception rate of the user was larger than that of the eavesdropper, the perfect secrecy was claimed to be achieved, where the estimation error covariance of the user is bounded while that of the eavesdropper grows unbounded, i.e.,
\begin{eqnarray}                              
  \begin{aligned}
  \label{perfect}
    \lim_{t\to\infty} \mathrm{Tr}(\mathbb{E}(P_{e}(t)))<\infty,\ 
    \lim_{t\to\infty} \mathrm{Tr}(\mathbb{E}(P_{u}(t)))=\infty,
  \end{aligned} 
\end{eqnarray}
where $P_{e}(t)$ is the estimation error covariance matrix of the eavesdropper and $P_{u}(t)$ is the that of the user. 
Wang et al. \cite{Wang_Drop} discussed the sensor scheduling problem under energy constraints, which maximized the estimation error covariance of the eavesdropper while maintaining that of the remote estimator at a certain level. 
The above scheduling problem was further transformed into three optimal problems and one energy allocation problem under arbitrary fixed transmission times.
Besides, the adversary in \cite{Lu_Drop} was assumed to overhear the transmissions and proactively hijack the sensor to reprogram its transmission policy. 
The trade-off between eavesdropping performance and detection risk was constructed by a constrained Markov decision process \cite{Ding_Drop}, and then it was changed into an unconstrained problem by resorting to a Lagrange multiplier. 
Also, under the energy constraint, the cost combining the estimation and privacy performances was minimized over the infinite time horizon in \cite{Huang_Drop}.

\subsection{Other Scheduling Methods} 
Recently, an effective encryption method called the state-secrecy code was proposed by Tsiamis \cite{Tsiamis_Code_CDC,Tsiamis_Code_ACC,Tsiamis_Code_TAC}. 
The reference time is denoted as $t_{k}$, and it represents the time slot at which the most recent successful reception occurred for the legitimate user.
Then, the state-secrecy coding scheme at time $t$ can be described as follows: 
\begin{eqnarray}                               
  \begin{aligned}
  \label{coding}
    z(t)=x(t)-A^{t-t_{k}}x(t_{k}).
  \end{aligned}
\end{eqnarray}
where $z(t)$ is the coding output utilized for transmission. 
The essence of this method hinges on a critical event, in which the legitimate user successfully acquires the transmitted package while the eavesdropper does not, i.e., 
\begin{eqnarray}                               
  \begin{aligned}
  \label{event}
    \gamma_{u}(t)=1,\ \gamma_{e}(t)=0. 
  \end{aligned}
\end{eqnarray}
If this event occurs for any time $t\in\mathbb{Z}_{+}$, the perfect secrecy can be achieved \cite{Tsiamis_Code_TAC}. 
The state-secrecy code for stable systems was proposed in \cite{Tsiamis_Code_ACC}, which imposed extra unstable dynamics on the recursive estimator of the eavesdropper. 
On the other hand, for unstable systems, the state transition matrix whose own dynamic was unstable was introduced to amplify the initial errors caused by the critical event \cite{Tsiamis_Code_TAC}. 
Meanwhile, the perfect secrecy defined in \eqref{perfect} was achieved, and the divergence rate of the eavesdropper's estimator approached that of the open loop. 
Besides, the coding scheme was also studied for distributed estimation \cite{Chen_Code}, where the communications in the wireless sensor networks were protected.

Nevertheless, the transmission of ACKs cannot be completely safe and it is also susceptible to cyberattacks \cite{Knorn_ACK,Li_ACK}. 
In the event that the sensor does not receive the ACK, the coding scheme depended on the most recent time when the legitimate user successfully received the package and the sensor also received the ACK \cite{Tsiamis_Code_TAC}. 
Moreover, Lücke et al. \cite{Lucke_Code} introduced a dynamic switching strategy designed to uphold resilience against flipped ACKs, which encompassed situations sending the ciphertexts and sending the plaintexts. 
In recent work, Kennedy et al. \cite{Kennedy_Code} proposed a state-secrecy coding scheme for remote state estimation without requiring ACKs. 
The method is based on randomly alternating between transmission of system state values and random noise with similar statistical properties as the system state. 
Through use of a pseudo-random sequence for pre-arranging the scheduling of the random noise, the legitimate user can discard noisy data, whereas the eavesdropper is deceived. 

Artificial noise is a typical physical-layer method about encrypting communication signals. 
The complex noise signals are injected into the modulated signals of original data to decrease the signal-to-noise ratio (SNR) of transmitted signals in the channels.
The degradation on SNR poses a challenge for the eavesdroppers to successfully demodulate the information, thus preventing them from acquiring any valid data. 
Since the general complex noises simultaneously affected the legitimate channels, a complex noise method based on null space design was proposed by Goel and Negi \cite{Goel_Noise}. 
According to this approach, the noise lies in the null space of the channel gain of the legitimate user, ensuring that the user can completely eliminate the extra inserted artificial noises. 
In this case, the finally transmitted signal $\mathrm{x}(t)$ can be expressed by the following form \cite{Goel_Noise}: 
\begin{eqnarray}                           
  \begin{aligned}
  \label{complex}
    \mathrm{x}(t)=\mathrm{s}(t)+\mathrm{w}(t),\ \mathrm{H}(t)\mathrm{w}(t)=0
  \end{aligned}. 
\end{eqnarray}
$\mathrm{s}(t)$ is signal vector of valid information and $\mathrm{H}(t)$ is the legitimate channel gain. 
$\mathrm{w}(t)=\mathrm{Z}(t)v(t)$, where $\mathrm{Z}(t)$ is an orthonormal basis for the null space of $\mathrm{H}(t)$ satisfying $\mathrm{Z}^{\dag}(t)\mathrm{Z}(t)=I$ and $v(t)$ is the complex Gaussian noise. 
Denote the channel gain of the eavesdropper as $\mathrm{G}(t)$. 
Then, on the basis of the above mechanism \eqref{complex}, the respective signals obtained for the legitimate user and that for the eavesdropper can be expressed as follows:
\begin{eqnarray}                           
  \begin{aligned}
  \label{signal}
    &z_{u}(t)=\mathrm{H}(t)\mathrm{s}(t)  \\
    &z_{e}(t)=\mathrm{G}(t)\mathrm{s}(t)+\mathrm{G}(t)\mathrm{w}(t)
  \end{aligned}. 
\end{eqnarray}
The result demonstrates that the legitimate user can effectively eliminate the added noises and extract the desired information, while the eavesdropper is unable to do so. 
Accordingly, when the noise is large enough, the eavesdropper cannot successfully demodulate the received signals because of the low SNR, which guarantees the perfect data privacy. 

Through adding noises to the transmitted information from sensors, Leong et al. \cite{Leong_Noise} realized the perfect secrecy and proposed the sufficient conditions for energy allocation. 
Xu et al. \cite{DaxingXu_Noise_ISA} designed a secure DFE with complex noises and simultaneously applied a dimensionality reduction method to reduce energy consumption.
A prediction-based compensation method was considered to mitigate the adverse impact generated by the dimensionality reduction.
Subsequently, by resorting to the event-trigger approach, the energy consumption of complex noises was significantly reduced, and the stability of the DFE was also guaranteed in \cite{DaxingXu_Noise_TCSII}. 
Furthermore, Xu et al. \cite{DaxingXu_Noise_TAC} discussed a similar DFE problem under energy constraints, where an optimization problem was established based on the terminal fusion covariance and the energy cost of artificial noises in the finite time horizon. 
The optimal scheduling sequence was proposed, which tried to encrypt the LSEs only during the last time period.

Moreover, Guo et al. \cite{Guo_Noise} constructed the power allocation problems to minimize the distortion outage probability for the FC, subject to a entire energy constraint and a secrecy outage constraint for the eavesdropper. 
They also studied two different scenarios \cite{Guo_Noise2}, including the single sensor case with multiple antennas and the multiple sensors case in which each sensor has a single antenna. 
The similar power allocation problems were solved, while the performances were compared varying from the number of sensors and antennas.
Besides, Yang et al. \cite{Yang_Code} applied the invertible linear transformation to encode the measurements with artificial noise, and then derived the lower bound of the such noise under a SNR threshold.

\begin{table*}[htbp]
\centering
\caption{Comparison of typical methods for Privacy-Preserving State Estimation.}
\begin{tabular}{|p{0.15\textwidth}<{\centering}|p{0.15\textwidth}<{\centering}|p{0.3\textwidth}<{\centering}|p{0.3\textwidth}<{\centering}|}
\hline
Method
& Some references 
& Advantages 
& Disadvantages \\
\hline
Homomorphic encryption, Secure multi-party computation
& \cite{Gonzalez_HE,Landa_HE,Aristov_HE,Ni_HE,Emad_HE,Ristic_HE,Mohsen_HE,Kim_HE,Alanwar_HE} 
& 1) The data accuracy can be maintained with secret keys; 
2) Certain operations can be done on ciphertexts. 
& 1) The computation amount is high; 
2) The distribution of secret keys needs to be completely safe. \\
\hline
Watermarking, pseudo-random samples
& \cite{Ristic_Pseudo,Huang_Water,Zhou_Water} 
& 1) The original data can be completely recovered; 
2) The computation amount is relatively low. 
& 1) The distribution of watermark should be entirely secure; 
2) The performance of the eavesdropper may not be thoroughly destroyed. \\
\hline
Differential privacy
& \cite{Ny_DP,Ny_DP2,Ny_DP_TAC,Ny_DP_Book,XinhaoYan_DP_TAES,Yazdani_DP,Degue_DP_Conf,Degue_DP,XinhaoYan_DP_Auto,Yan_DP}
& 1) The encryption procedure is simple; 
2) The mathematical model is rigorous, and the privacy level can be adjusted varying from the practical demands.
& 1) The information is still leaked on a certain level; 
2) The assumption on adjacent relation is not practical enough. \\
\hline
Null space perturbation
& \cite{XinhaoYan_Eaves_TAES,XinhaoYan_DP_Auto} 
& 1) Some components are stealthily eliminated while requiring a few computations; 
2) The performance of the eavesdropper can be damaged even for stable systems. 
& 1) The sensors need to communicate with each other; 
2) The assumption for the eavesdropper is too strong and not practical. \\
\hline
Optimal scheduling
& \cite{Aysal_Drop,Reboredo_Drop,Yuan_Drop,Leong_Drop_LCSYS,Leong_Drop_arXiv,Leong_Drop_CDC,Tsiamis_Drop_IFAC,Wang_Drop,Lu_Drop,Ding_Drop,Huang_Drop} 
& 1) The assumption for the eavesdropper is weak, and it can be applied to most systems; 
2) Optimal expected target can be achieved. 
& 1) The estimation error of the eavesdropper cannot diverge. 
2) The assumption on random packet drop link may be outdated due to the development of communication technology. \\
\hline
State-secrecy code
& \cite{Tsiamis_Code_CDC,Tsiamis_Code_ACC,Tsiamis_Code_TAC,Lucke_Code,Kennedy_Code} 
& 1) The legitimate user can decode to acquire the original data; 
2) The performance of the eavesdropper approaches the open loop. 
& 1) The MSE of the eavesdropper cannot diverge for stable systems. 
2) The time of successful packet reception should be known. \\
\hline
Artificial noise
& \cite{Goel_Noise,Leong_Noise,DaxingXu_Noise_ISA,DaxingXu_Noise_TCSII,DaxingXu_Noise_TAC,Guo_Noise,Guo_Noise2,Yang_Code}
& 1) The eavesdropper will lose most packets; 
2) The legitimate user can get all the information. 
& 1) Extra energy will be consumed; 
2) It is unsuitable for time-varying channels. \\
\hline
\end{tabular}
\label{Table_Methods}
\end{table*}

\subsection{Discussion} 
When compared with the system constructions of other methods, the assumption about random packet drop link is more practical because of the channel modeling. 
The trade-off optimization methods based on the performances of both user and eavesdropper can be applicable for most systems, since there is no strong hypothesis for the eavesdropper. 
By contrast, artificial noise is quite effective on protecting high transmission privacy without considering the system stability. 
When the noise is large enough, the eavesdropper cannot acquire any valid information, because it cannot successfully demodulate the perturbed signals, whose form is similar to the unsuccessful decryption with cryptography. 

Since the eavesdropping attack is a kind of passive and covert attack, we cannot accurately speculate the position or other information of the potential eavesdroppers. 
This means that the stealthily established communication channels of eavesdroppers are random in a certain range and unknown to legitimate systems. 
Therefore, the packet drop link assumption for eavesdroppers may be impractical sometimes, because it is challenging to determine the packet drop probabilities of the eavesdroppers' channels. 
In this case, the construction of null space-based artificial noise may not be successful, and thus its effectiveness cannot be guaranteed. 

On the other hand, the artificially complex noises will consume a lot of energy, because the energy should be large such that the SNR is low enough to ensure the sufficient confidentiality. 
However, a practical consideration is that in many systems, sensors are powered by the batteries that have a limited energy supply. 
In this case, the employment of artificial noise may decrease the normal working hours of the sensors. 
Besides, the state-secrecy code approach just realizes the divergence of the eavesdroppers' estimators for unstable state components, while the eavesdropper can still obtain the information about bounded ones.

\section{Conclusion}
This paper offers a comprehensive overview of privacy-preserving approaches in the context of state estimation systems. 
It specifically focuses on three main encryption methods, including cryptography, data perturbation, and transmission scheduling. 
We extensively discuss the fundamental models, classical applications, and recent developments related to these encryption approaches. 
Especially, we explore various privacy-preserving estimators while thoroughly evaluating their strengths and weaknesses. 
Every method has its benefits and drawbacks. 
It is inappropriate to say for sure which method is better or worse without knowing the specific scenario.
The comparison of some typical methods is given in Table \ref{Table_Methods}. 

In what follows, we give some recommendations for future research based on this survey: 

\begin{enumerate}
\item \emph{Computational-Efficient Cryptography}: 
Most cryptographic methods such as Homomorphic encryption require high computation amounts. 
They may affect normal operations, while the real-time performance is quite important for estimation systems. 
Thus, the encryption structure can be simplified to reduce extra computation, for example, using partial encryption to process distributed state estimation. 
\item \emph{Energy-Efficient Noise Injection}: 
The generation of artificial noises requires a great deal of power, while the total energy of the system is always finite. 
Meanwhile, the channels are time-varying in many scenarios such as maneuvering vehicles, which renders the method unsuitable or ineffective. 
Hence, constructing some optimization problems to reduce the influence of such unexpected terms is quite valuable. 
\item \emph{Low-impact Perturbation}: 
The general perturbation processes always affect the legitimate performance while protecting privacy, and there exist some tradeoffs \cite{Li_Tradeoff}. 
Some approaches such as null space \cite{XinhaoYan_Eaves_TAES} and coding \cite{Tsiamis_Code_TAC} can be utilized to eliminate the adverse impact. 
However, the necessary conditions of system parameters may be difficult, and thus some relaxation or suboptimal cases can be further studied. 
\item \emph{Multiple Attacks}: 
In practical applications, it is common for systems to be subject to multiple attacks simultaneously \cite{Yao_Multiple}. 
The direct combination of various individual defense approaches will increase the complexity and cost of legitimate systems. 
Alternatively, countering multiple attacks with only one single scheme must be more efficient. 
Furthermore, it is also valuable to apply the effect of one attack to impact another \cite{Yuan_Drop}. 
\item \emph{Physical Layer-Based Approach}: 
The information obtained from the physical layer plays a crucial role and carries unique significance. 
It can be utilized to add in the encryption \cite{DaxingXu_Noise_ISA}, and can even be regarded as the secret keys that are difficult for eavesdroppers to acquire. 
In this case, appropriate usage of the physical-layer data can enhance the effectiveness of the methods \cite{Zhou_CPS}. 
\item \emph{Artificial Intelligence-Based Approach}: 
Artificial intelligence approaches such as neural networks have been utilized for defending cyberattacks \cite{Farivar_AI}, while there is few related study on confidential estimation method. 
According to the ideas in \cite{Farivar_AI}, artificial intelligence methods show promise in the development of efficient encryption or compensation techniques. 
\end{enumerate}

\vfill
\end{document}